\documentclass[conference, a4paper]{IEEEtran}

\usepackage[utf8]{inputenc} 
\usepackage[T1]{fontenc}
\usepackage{url, balance}              

\usepackage[cmex10]{amsmath}  
\usepackage{amsthm}
\usepackage{amsfonts}
\usepackage{amssymb}
\usepackage{thmtools}
\usepackage{thm-restate}
\usepackage{graphicx}
\usepackage{nicefrac}
\usepackage{bm}
\usepackage{xcolor, xpatch}
\usepackage[linesnumbered,ruled]{algorithm2e}
\usepackage{tikz}
\usetikzlibrary{positioning, arrows.meta}
\usepackage{enumitem}
\usepackage{booktabs}         
\usepackage[backend=bibtex,bibstyle=ieee,citestyle=numeric-comp,url=false,doi=false,isbn=false]{biblatex}

\usepackage[draft]{hyperref}

\usepackage{cleveref}
\usepackage{pgfplots}

\makeatletter
\xpatchcmd{\proof}{\topsep6\p@\@plus6\p@\relax}{}{}{}
\makeatother

\newtheorem{thm}{Theorem}[section]
\newtheorem{prop}[thm]{Proposition}

\newtheorem{remark}[thm]{Remark}
\newtheorem{cor}[thm]{Corollary}

\newtheorem{defi}[thm]{Definition}

\newcommand{\bfc}{{\boldsymbol c}}

\newcommand{\bfs}{{\boldsymbol s}}

\newcommand{\bfu}{{\boldsymbol u}}

\newcommand{\bfx}{{\boldsymbol x}}
\newcommand{\bfy}{{\boldsymbol y}}

\newcommand{\cC}{\mathcal{C}}

\newcommand{\cL}{\mathcal{L}}

\IEEEoverridecommandlockouts

\newcommand{\Fq}{\mathbb{F}_q}

\newcommand{\dham}{d_{H}}
\newcommand{\ded}{d_{\textup{ed}}}
\newcommand{\balpha}{{\boldsymbol\alpha}}

\newcommand{\ex}{\mathbb{E}}

\newcommand{\pin}{P_{\mathrm{i}}}
\newcommand{\pd}{P_{\mathrm{d}}}

\newcommand{\pt}{P_{\mathrm{t}}}

\newcommand{\floor}[1]{\lfloor #1 \rfloor}

\bibliography{resources/literature.bib}

\begin{document}
\title{Decoding Insertions/Deletions via List Recovery \\

\thanks{The research was Funded by the European Union (ERC, DNAStorage, 101045114 and EIC, DiDAX 101115134). Views and opinions expressed are however those of the authors only and do not necessarily reflect those of the European Union or the European Research Council Executive Agency. Neither the European Union nor the granting authority can be held responsible for them.

This work was also supported by a DFG Middle-East Grant, under grant number WA3907/12-1. 
}

}

\author{%
  \IEEEauthorblockN{Anisha Banerjee\IEEEauthorrefmark{1},  Roni Con\IEEEauthorrefmark{2},
                    Antonia Wachter-Zeh\IEEEauthorrefmark{1},
                    and Eitan Yaakobi\IEEEauthorrefmark{2}}
  \IEEEauthorblockA{\IEEEauthorrefmark{1}%
                    Institute for Communications Engineering, Technical University of Munich (TUM), Munich, Germany}
  \IEEEauthorblockA{\IEEEauthorrefmark{2}%
                    Department of Computer Science, Technion - Israel  Institute  of  Technology, Haifa 3200003, Israel}
    \IEEEauthorblockA{Email: anisha.banerjee@tum.de, 
      roni.con93@gmail.com, 
      antonia.wachter-zeh@tum.de,
      yaakobi@cs.technion.ac.il}
      \\[-3.6ex]

}
\maketitle

\begin{abstract}
    In this work, we consider the problem of efficient decoding of codes from insertions and deletions.
    Most of the known efficient codes are codes with synchronization strings which allow one to reduce the problem of decoding insertions and deletions to that of decoding substitution and erasures.
    Our new approach, presented in this paper, reduces the problem of decoding insertions and deletions to that of list recovery.
    Specifically, any \((\rho, 2\rho n + 1, L)\)-list-recoverable code is a \((\rho, L)\)-list decodable insdel code. As an example, we apply this technique to Reed–Solomon (RS) codes, which are known to have efficient list-recovery algorithms up to the Johnson bound. 
    In the adversarial insdel model, this provides efficient (list) decoding from \(t\) insdel errors, assuming that \(t\cdot k = O(n)\). This is the first efficient insdel decoder for \([n, k]\) RS codes for \(k>2\). Additionally, we explore random insdel models, such as the Davey-MacKay channel, and show that for certain choices of \(\rho\), a \((\rho, n^{1/2+0.001}, L)\)-list-recoverable code of length \(n\) can, with high probability, efficiently list decode the channel output, ensuring that the transmitted codeword is in the output list. In the context of RS codes, this leads to a better rate-error tradeoff for these channels compared to the adversarial case. We also adapt the Koetter-Vardy algorithm, a famous soft-decision list decoding technique for RS codes, to correct insertions and deletions induced by the Davey-MacKay channel.

\end{abstract}

\section{Introduction} \label{sec::intro}
Error-correcting codes (codes, for short) are designed to enable the recovery of original information from data that has been corrupted. 
The primary corruption models studied are substitutions or erasures. In these models, each symbol in the transmitted word is either replaced with a different symbol from the alphabet (a substitution) or with a '?' (an erasure).

Another important model that has been considered since the works of Shannon and Hamming is that of \emph{insertions and deletions} (insdel errors, for short). An insertion error is when a new symbol is inserted between two symbols of the transmitted word and a deletion is when a symbol is removed from the transmitted word. 
For example, over the binary alphabet, when $100110$ is transmitted, we may receive the word $1101100$, which is obtained from two insertions ($1$ at the beginning and $0$ at the end) and one deletion (one of the $0$'s at the beginning of the transmitted word). 
The insdel error model is as natural as that of substitutions and erasures; however, with the loss in synchronization between the sender and the receiver, insdel errors are much more challenging to handle.

Motivated by applications such as over-/under-samping and DNA-based data storage, in recent years many researchers studied and designed codes that can correct from insdel errors 
\cite{schulman1999asymptotically,levenshtein2002bounds,haeupler2021synchronization-org,brakensiek2017efficient,guruswami2017deletion,schoeny2017codes,gabrys2018codes,cheng2018deterministic,sima-q2020optimal,sima2020systematic,haeupler2019optimal,cheng2020efficient,haeupler2021synchronization,guruswami2020optimally,guruswami2021explicit,guruswami2022zero}, just to name a few. These works, among many others, contributed a lot to our knowledge about dealing with insdel errors. However, still our understanding of this model lags far behind our understanding of codes that correct erasures and substitution errors (we refer the reader to the following excellent surveys \cite{mitzenmacher2009survey,cheraghchi2020overview,haeupler2021synchronization}).


In this paper, we address the problem of decoding codes against insdel errors. We first describe the following simple coding scheme that combines classical codes (capable of correcting errors and erasures) and indices. Specifically, for any codeword $\bfc = (c_1, \ldots, c_n)\in \cC$, we create a new codeword $\bfc' = ((i_1,c_1), \ldots, (i_n,c_n)) \in \cC'$ where we observe that the alphabet of the new code is larger (by a factor of $n$).
At the decoding side, upon receiving \[((i_1, y_{i_1}), \ldots, (i_m,y_{i_m})) \;, \] the decoder first decodes the indices using the following simple rule. For every $j\in [n]$, if $j$ does not appear in $i_1, \ldots, i_m$ or if $j$ appears more than once, declare $\widetilde{\bfc}_j$ to be an erasure. Otherwise, $j = i_s$ for some unique $s\in [m]$, and set $\widetilde{\bfc}_j = y_{i_s}$.
It is easy to see that if $\delta n$ insdel errors occur to $\bfc'$, then $\widetilde{\bfc}$ can be obtained from $\bfc$ by performing $t$ substitutions and $e$ erasures.

There is a main drawback with this construction. The alphabet size of the code $\cC'$  grows with $n$.\footnote{In fact, in order to make the rate of the code $\cC'$ close to $\cC$, the alphabet of $\cC$ needs to be at least $\Omega(n^{1/\varepsilon})$. } 
In a breakthrough work \cite{haeupler2021synchronization-org,haeupler2018synchronization}, Haeupler and Shahrasbi constructed a sequence of indices $(i_1, \ldots, i_n)$, termed as \emph{$\varepsilon$ self-matching string}, over finite alphabet size such that if we replace the indices $(1, \ldots, n)$ with $(i_1, \ldots, i_n)$ in the construction of $\cC'$ then we get that $\cC'$ is a code over an alphabet of size $O_{\varepsilon}(1)$ that can correct from $\delta n$ insdel errors and has a rate of $1 - \delta - \varepsilon$.

In this paper, we present a different approach for decoding codes against insdel errors. 
Our idea is to reduce the problem of insdel decoding to that of \emph{list recovery}. 
Namely, if $\cC$ is a list-recoverable code with an efficient list recovery algorithm
, then, we can efficiently \emph{list decode} $\cC$ from insdel errors. The number of insdel errors we can list decode from is determined by the parameters of the list recoverable codes. See \Cref{thm:adv-decode} for the formal statements.
This approach allows us to efficiently decode insdel errors in several classes of codes. As an example for our approach, we consider the well-known Reed--Solomon (RS, for short) codes which have an efficient list-recovery algorithm up to the \emph{Johnson bound} \cite{guruswami1998improved} and get an efficient unique decoding algorithm for RS codes against insdel errors (the amount of insdel errors that can be corrected efficiently is determined by the Johnson bound).
Although RS codes have been studied in the context of insdel errors in recent years \cite{safavi2002traitor,wang2004deletion,tonien2007construction,duc2019explicit,liu20212,con2023reed,con2024optimal,liu2024optimal,con2024random}, the only efficient non-trivial decoding algorithm that decodes RS codes from insdel errors was given in \cite{singhvi2024optimally}.
Specifically, in \cite{singhvi2024optimally}, the author presents a decoding algorithm for the $[n,2]_q$ RS code constructed in \cite{con2024optimal} that can decode up to $n-3$ deletions in linear time.

As a consequence of our results, we obtain a polynomial time decoding algorithm for $[n,k]_q$ RS codes that can correct up to $t$ insdel errors as long as $t\cdot k =O(n)$.

Subsequently, we also adapt this analysis to the Davey-MacKay channel \cite{daveyReliableCommunicationChannels2001}, a probabilistic insdel channel. Therein, a better rate-error tradeoff is observed. Finally, we modified a well-known soft-decision list decoding algorithm for Reed-Solomon (RS) codes, namely the Koetter-Vardy algorithm~\cite{koetterAlgebraicSoftdecisionDecoding2003}, to make it compatible with this channel. 

\subsection{Preliminaries}

Throughout, $\Fq$ will denote a finite field of order $q$.  
We denote by $d_H(\cdot, \cdot)$, the Hamming distance between two strings. 
We shall use the notation $\binom{\Fq}{\ell}$ to refer to all subsets of size $\ell$ of $\Fq$. Further, we write  $\binom{\Fq}{\ell}^n$ to refer to all possible $S_1 \times S_2 \times \cdots \times S_n$ where each $S_i$ is a subset of $\Fq$ of size $\ell$.
Finally, we write $\dham(c, S):=\left| \{i\in [n] | c_i \notin S_i\} \right|$.
With these notations, we are ready to define a \emph{list recoverable code}.
\begin{defi}
	For $\rho \in [0,1]$ and integers $\ell, L$, we say that a code $\cC \subseteq \Fq^n$ is $(\rho, \ell, L)$-\emph{list-recoverable} if for any $S\in \binom{\Fq}{\ell}^n$ there are at most $L$ different codewords $c \in \cC$ which satisfy that $\dham(c, S)\leq \rho n$.
\end{defi}

List recoverable codes are an important class of codes that have found applications in domains such as pseudorandomness~\cite{guruswami2009unbalanced}, algorithms~\cite{doron2022high}, combinatorial group testing~\cite{indyk2010efficiently}, hashing, and many more. As such, studying the limitation and constructing (efficient) list recoverable codes is a very active area of research \cite{rudra2018average,lund2020list,tamo2024tighter,guo2024improved,kopparty2020list}, just to name a few.
The Johnson bound for list recoverable codes (see e.g., \cite[Theorem 2.3]{lund2020list}) states that a code $\cC$ with relative Hamming distance $\delta$ is $(\rho, \ell, L)$ for any $\ell <  \frac{(1 - \rho)^2}{1 - \delta}$ with $L = \frac{\ell}{(1 - \rho)^2 - (1 - \delta)\ell}$.
Observe that a $(\rho, 1, L)$-list-recoverable code is a list decodable code. Thus, list-recoverable codes are a generalization of the more widely studied list-decoding model.

Recall that we are dealing with insdel errors in this paper, and the metric that we are going to use is the \emph{edit} metric.
\begin{defi}
        The \emph{edit distance} between $\bfs$ and $\bfs'$, denoted by $\ded(\bfs,\bfs')$, is the minimal number of insertions and deletions needed in order to turn $\bfs$ into $\bfs'$. 
\end{defi}
For a code $\cC$ we write, $\ded (\cC) = \max_{\bfc, \bfc'\in \cC, \bfc \neq \bfc'} \ded(\bfc, \bfc')$.
Note that list decoding can also be defined in the insdel model. Namely, we say that $\cC$ is $(\rho, L)$ \emph{list decodable insdel code} if for every $\bfy$, it holds that $|\{\bfc \in \cC\ |\ \ded(\bfc, y)\leq \rho n \}|\leq L$.

We now formally define RS codes.
\begin{defi}
    Let $\balpha = (\alpha_1, \alpha_2, \ldots, \alpha_n) \in \Fq^n$ where $\alpha_i \neq \alpha_j$ for every distinct $i,j\in [n]$. The $[n,k]_q$ \emph{Reed--Solomon} (RS) code of dimension $k$ and block length $n$ associated with the evaluation vector $\balpha$ is defined to be the set of codewords 
    \[
    \text{RS}_{n,k}(\balpha):= \left \lbrace \left( f(\alpha_1), \ldots, f(\alpha_n) \right) \mid f\in \mathbb{F}_q[x],\text{ }\deg f < k \right \rbrace.
    \]
\end{defi}

RS codes are among one of the most widely used codes in theory and practice. They have efficient unique decoding algorithms up to half of the minimum distance and also efficient list decoding algorithms up to the Johnson bound~\cite{guruswami1998improved}. Note here that since RS codes have $R = 1 - \delta$, then we have that $\rho < 1- \sqrt{\ell R}$.

\subsection{Our Results}
In this paper, we present a connection between list recoverable codes and insdel codes. This connection is achieved by a simple reduction where we take an input to an insdel decoding algorithm and transform it to an input to a list recovery algorithm. This implies that if one has a code with an efficient list recovery algorithm, then the code has also an efficient insdel decoding algorithm. 

Then, we also apply our idea to probabilistic errors and show that over random insdel channels we can slightly improve the rate-error correction trade-off. 

More specifically, we prove the following statements.
\begin{itemize}
    \item Any $(\rho, 2\rho n + 1, L)$-list-recoverable code $\cC$ is an $(\rho, L)$-list decodable insdel code. Moreover, if $\rho n$ is smaller than half the edit distance of $\cC$, then the decoding is unique (see \Cref{thm:adv-decode} for a formal description).
    \item Let $\pin, \pd\in (0,1)$ and $\varepsilon > 0$. Let $\cC$ be a $(E + \varepsilon, O(n^{1/2 + 0.01}), L)$-list-recoverable code, where $E$ is a constant depending on $\pin$ and $\pd$.
    Then, $\cC$ is a list decodable code in the random insdel model (the Davey-MacKay channel \cite{daveyReliableCommunicationChannels2001} with a constraint on the maximum length of an insertion burst) with insertion and deletion probabilities being $\pin$ and $\pd$ respectively. Namely, with high probability, upon transmitting any codeword $\bfc\in \cC$ through this channel, we can efficiently list decode the output so that $\bfc$ is in the output list (see \Cref{thm::davey-ch} for the formal statements). An analogous result for the probabilistic deletion channel is also presented in \Cref{thm::del-ch}.
\end{itemize}

Finally, we adapt the Koetter-Vardy algorithm \cite{koetterAlgebraicSoftdecisionDecoding2003} to insdel channels. This is the only decoding algorithm for RS codes that can leverage soft information. Additionally, we show how to jointly decode multiple received sequences, which is useful in applications like DNA data storage that provide multiple erroneous copies~\cite{churchNextGenerationDigitalInformation2012, goldmanPracticalHighCapacityLowMaintenance2013, maaroufConcatenatedCodesMultiple2023, welterIndexBasedConcatenatedCodes2023, sakogawaSymbolwiseMAPEstimation2020, srinivasavaradhanTrellisBMACoded2021a}. 
We also present simulation results for decoding insdel errors in RS codes for various regimes.

\section{Decoding adversarial insdels via list-recovery}
This section shows our main result about decoding insdel errors via list recovery. We then consider the specific case of RS codes with an efficient list recoverable algorithm.

\begin{algorithm}[htb]	
	\caption{
	$\mathtt{Decode}$ 
	}
	\label{alg:decode-adv}
    \DontPrintSemicolon
	\KwIn{$y_1, \ldots, y_m$, odd integer $\ell$} 
	\KwOut{A codeword $c \in \cC$}
        \For{$i\in [n]$}{$S_i \gets \{ y_{\max \{1, i - \floor{\ell/2}\}},\ldots, y_{\min\{n, i+\floor{\ell/2}\}} \} $}
        Execute $\mathtt{List}$-$\mathtt{Recover}$ with the lists $S_1, \ldots, S_n$ to get $\cL \subseteq \cC$. \label{alg:decoce-list-recovery}\\
        return $\{ \bfc \in \cL | \ded(\bfc, \bfy) \leq \rho n \}$.
	\BlankLine
\end{algorithm}
\begin{thm} \label{thm:adv-decode}
	Let $\rho \in [0,1]$ and set $\ell = 2\rho n + 1$. \footnote{for the sake of notations we assume that $\rho n$ is an integer.}
	Assume that $\cC$ is a $(\rho, \ell, L)$-list-recoverable code with an algorithm $\mathtt{List}$-$\mathtt{Recover}$ that runs in time $T$. 
	Let $\bfy \in \Fq^m$. Then, running Algorithm~\ref{alg:decode-adv} on input $\bfy$ returns a list $\cL'$ of size $|\cL'| \leq L$ such that 
	for every $\bfc \in \cL'$, we have $\ded(\bfc, \bfy) \leq \rho n$. The running time of Algorithm~\ref{alg:decode-adv} is $O(T + L\cdot n^2)$.
    
	Moreover, if $\rho n \leq \floor{\frac{\ded(\cC) - 1}2}$, then $|\cL'| \leq 1$.

\end{thm}

\begin{IEEEproof}
	First note that by the definition of the $S_i$'s, we have that $|S_i|\leq \ell$ for all $i\in [n]$. 
	We now show that for every $\bfc \in \cC$, such that $\ded (\bfc, \bfy) \leq \rho n$, we have that $\dham(\bfc, S)\leq \rho n$.
	Indeed, denote by $a$ and $b$ the number of deletions and insertions, respectively, that are needed to transform $\bfc$ to $\bfy$.
	
	Let $c_i$ be the $i$th symbol of $\bfc$ and assume that it was not deleted by any of the $a$ deletions and let $j$ be its position in $\bfy$.
	We have that $-a \leq j - i\leq b$ and since $a$ and $b$ are at most $\rho n$, we must have that $c_i \in S_i$.
	Thus, the number of $i$s for which $c_i \notin S_i$ is at most $a \leq \rho n$ and we conclude that $c\in \cL$.
	
	Finally, the list-recovery algorithm returns a list $\cL$ of size at most $L$. Thus, in the last step of the algorithm, we can only reduce the set $\cL$ to obtain $\cL'$. 
	
	The running time follows by observing that setting the $S_i$s takes $O(n^2)$, and the last step takes $O(L \cdot n^2)$ since computing the edit distance between two strings takes $O(n^2)$ time.


    To prove the moreover part, assume that $\rho n \leq\floor{\frac{\ded(\cC) -1}{2}}$ and assume that $\cL'$ contains at least two codewords, $\bfc$ and $\bfc'$. Then, by the triangle inequality, 
    $\ded(\bfc, \bfc')\leq \ded(\bfc, \bfy) + \ded(\bfc', \bfy) \leq 2\cdot\floor{\frac{\ded(\cC) -1}{2}} <  \ded(\cC)$, which is a contradiction.
\end{IEEEproof}

\begin{remark}
    We observe that if one needs to correct only from deletions or only from insertions, then we can set $\ell = \alpha n$ instead of $2\alpha n +1$ in \Cref{thm:adv-decode}.
    
\end{remark}

We can apply the above theorem with list-recoverable codes. For RS codes, we apply the polynomial time Guruswami-Sudan decoder \cite{DBLP:conf/focs/GuruswamiS98}, and obtain the following corollary. 
\begin{cor} \label{cor:rs-unique-decode}
    Let $\varepsilon > 0$ and let $\cC$ be an $[n,k]_q$ RS code that can correct from $t$ insdel errors where 
    \begin{equation} \label{eq:insdel-rate-error}
        t \leq n - \sqrt{(1 + \varepsilon)\cdot kn \cdot (2t + 1)}\;.
    \end{equation}
    Then, there is a deterministic insdel unique-decoding algorithm for $\cC$ that can correct $t$ insdel errors in time $O(n^3\varepsilon^{-6})$.
\end{cor}

\begin{remark} \label{rem:bad-rate-error-adv}
    Observe that for the inequality \eqref{eq:insdel-rate-error} in \Cref{cor:rs-unique-decode} to hold, we cannot have $t = \Omega(n)$ and $k = \Omega(n)$. In fact, it must be that $k\cdot t = O(n)$.
\end{remark}

\section{Probabilistic Channels}

\begin{figure}[htb]
	\centering
	\begin{tikzpicture}[->,>=stealth,auto,node distance=3.5cm,scale=0.6, every node/.style={scale=0.9}]
	\node[name=a, circle, inner sep=0pt, draw=black,fill=white, minimum size=1cm] at (0,-2){$x_{i}$};
	\node[name=b, circle, inner sep=0pt, draw=black,fill=white, minimum size=1cm] at (8.0,-2){$x_{i+1}$};
	\node[anchor=east,xshift=-15pt] at (a) {$\cdots$};
	\node[anchor=east,xshift=35pt] at (b) {$\cdots$};
	\path[]
	(a) edge[bend right=30, dashed] node [midway, below, yshift=-.7mm] {$\pd$} node[midway, above] {Delete}(b)
	(a) edge[bend left=30] node [midway, above, right, yshift=3mm, xshift=-8mm,name=tr] {Transmit} node[midway, below] {$\pt$} (b);
	\draw (a) edge[loop above, out=120, in=60,min distance=20mm] node {Insert} node[midway, left, xshift=-5mm, yshift=-8mm] {$\pin$}(a);
	\end{tikzpicture}
	\caption{Allowed transitions in the insertion and deletion channel \cite{daveyReliableCommunicationChannels2001}}
	\label{fig::si}
\end{figure}
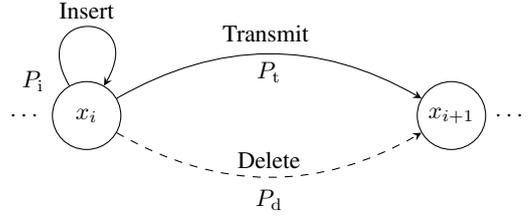

This section seeks to extend the application of Algorithm~\ref{alg:decode-adv} to probabilistic channels that corrupt the transmitted sequence with \emph{random} insertion and deletion errors. 
We focus on the Davey-MacKay channel \cite{daveyReliableCommunicationChannels2001}, which is parameterized by its insertion and deletion probabilities, denoted by $\pin$ and $\pd$, respectively. It describes a channel that involves insertion, deletion and substitution errors, as a finite state machine. In this work, substitution errors are ignored and Fig.~\ref{fig::si}, which illustrates this simplified channel model, suggests that for each symbol that awaits transmission, say $x_m$, one of three events may occur: a random symbol is inserted into the received stream with probability $\pin$ and $x_m$ remains in the transmission queue; or the next bit queued for transmission, i.e., $x_m$, is deleted with probability $\pd$; or $x_m$ is received with probability $\pt=1-\pin-\pd$.

 A significant challenge posed by such channels is the loss of synchronization in symbol positions. To quantify the extent of misalignment between the transmitted and the received symbols, we use the concept of the \emph{drift} \cite{daveyReliableCommunicationChannels2001}. 
\begin{defi}
    The integer drift $D_m$ is the difference between the number of insertions and the number of deletions that occurred until the $m$th bit has been queued for transmission.
\end{defi}


\subsection{The Deletion Channel} \label{subsec::del-ch}

We begin by examining a specific case of the Davey-MacKay channel that does not produce insertion errors, i.e., $\pin=0$. This is commonly known as the deletion channel, wherein each transmitted symbol is deleted independently with probability $\pd \in (0,1)$. By utilizing the statistical properties of the drift variable $D_m$, we establish the following bound concerning the error probability of a list-recovery decoder that uses sufficiently large lists for each transmitted symbol.

\begin{prop} \label{prop::del-ch}
    Assume a sequence $\bfx \in \Fq^n$ is transmitted over a channel with deletion probability $\pd$, resulting in the received sequence $\bfy \in \Fq^r$, where $r < n$. If for all $m \in [n]$, we assign $S_m \xleftarrow{} \{y_{a}, y_{a+1}, \ldots, y_{b} \}$, where $a = \max(m + \ex[D_m] - C, 1)$ and $b = \min(m + \ex[D_m] + C - 1, r)$, where $C$ is chosen such that $|S_m| \leq 2C = n^{\frac{1}{2} + 0.001}$ and let $I = \{ i | x_i \not\in S_i \}$, then it holds that
    \begin{equation*}
        \Pr[|I| \geq (P_d + \varepsilon) n] \leq \exp(-\Omega(n^{0.002})).
    \end{equation*}
\end{prop}
\begin{IEEEproof}
We begin by observing that for any $m\in [\varepsilon n, n]$, $\Pr[x_m \not\in S_m | x_m \textup{ was transmitted}]= \Pr[|D_m - \ex[D_m]| > C]$. Since $D_m$ can be seen as a sum of $m$ random variables, each iid and bounded within $[-1,0]$, we apply Hoeffding's inequality \cite{mitzenmacherProbabilityComputingRandomized2005} and the fact that $m \geq \varepsilon n$ to arrive at
\begin{IEEEeqnarray*}{+rCl+x*}
    \Pr[x_m \not\in S_m | x_m \textup{ was transmitted}]&\leq&  2 e^{-{2C^2}/{m}} \leq e^{-\Omega(n^{0.002})}.
\end{IEEEeqnarray*}

By the union bound, we further deduce that the probability that there exists an $i\in [\varepsilon n, n]$ such that $x_i$ was transmitted and $x_i\notin S_i$ is at most $n \cdot \exp(-\Omega(n^{0.002})) = \exp(-\Omega(n^{0.002}))$.

Also note that since the event that any $x_i$ is transmitted, can be modeled by a Bernoulli random variable with success probability $1-\pd$, it follows from Chernoff's bound \cite{mitzenmacherProbabilityComputingRandomized2005} that $\Pr[|\bfy| < (1 - \pd - \varepsilon) n] \leq \exp(-\Omega(n))$.

Thus, to conclude, we bound the probability that $|I|\geq (\pd + \varepsilon)n$ by the sum of the probability that $|\bfy| < (1 - \pd - \varepsilon) n$ and the probability that given that $|\bfy| \geq (1 - \pd - \varepsilon)n$, there exists an $i\in [n]$ such that $x_i$ was transmitted and $x_i \notin S_i$. 
The proposition follows.
\end{IEEEproof}

\Cref{prop::del-ch} suggests that any list-recoverable code can be used over deletion channels to correct all errors with very high probability. This is stated more formally as follows.

\begin{thm} \label{thm::del-ch}
    Let $\pd\in (0,1)$ and let $\varepsilon > 0$.
    Let $\cC$ be an $[n,k]_q$ code that is $(\pd + \varepsilon, n^{1/2 + 0.001}, L)$-list-recoverable with an efficient list recoverable algorithm $\mathtt{List}$-$\mathtt{recover}$. 
    Let $\bfc\in \cC$ be any codeword and denote by $\bfy$ the output of the deletion channel with probability $\pd$ when given $\bfc$ as an input.
    
    Let $S_1, \ldots, S_n$ be the lists generated from $\bfy$ according to \Cref{prop::del-ch}. Then, with probability $\exp(-\Omega(n^{0.002}))$, running $\mathtt{List}$-$\mathtt{recover}$ produces a list $\cL$ such that $\bfc \in \cL$. 

    Moreover, if $(\pd + \varepsilon)n \leq \floor{\frac{\ded(\cC) - 1}{2}}$, then $|\cL| = 1$.
\end{thm}

\begin{remark} \label{rem::del-rate-tradeoff}
    Note that in contrast to \Cref{rem:bad-rate-error-adv}, here, as one would expect, the rate-error tradeoff is better. 
    Indeed, for the unique decoding case (i.e., the ``moreover'' part of \Cref{thm::del-ch}), in order to decode (with high probability) from any constant deletion probability $\pd \in(0,1)$, we need that $k = O(n^{1/2 - 0.001})$.
\end{remark}


\subsection{The Davey-MacKay Channel} \label{sec::davey}


We now endeavor to generalize \Cref{thm::del-ch} to channels that also permit insertions. Since the length of an insertion burst has a geometric distribution, as suggested by Fig.~\ref{fig::si}, we assume as in \cite{daveyReliableCommunicationChannels2001}, for ease of analysis, that any burst of insertions is limited to $B$ symbols. This implies that
\begin{IEEEeqnarray*}{+rCl+x*}
    \Pr[D_1 = i] &=& \begin{cases}
                    \delta_{B} \pd & \text{if } i = -1,  \\
                    \delta_{B} (1-\pin)(1-\pd) \pin^i & \text{if } i \in [0,B],
                 \end{cases}
\end{IEEEeqnarray*}
where $\delta_B = \pd+(1-\pd)(1-\pin^{B+1})$ is a normalizing constant. Now to establish a result analogous to \Cref{prop::del-ch}, we exploit the statistical properties of \( D_m \) as before. 

\begin{prop} \label{prop::davey-ch}
    Assume a Davey-MacKay channel with insertion and deletion probabilities $\pin$ and $\pd$ respectively, and the maximum number of consecutive insertions limited to a finite value. Consider a sequence $\bfx \in \Fq^n$ that is transmitted over this channel, resulting in the received sequence $\bfy \in \Fq^r$. If for all $m \in [n]$, we assign $S_m \xleftarrow{} \{y_{a}, y_{a+1}, \ldots, y_{b} \}$, where $a = \max(m + \ex[D_m] - C, 1)$ and $b = \min(m + \ex[D_m] + C - 1, r)$, where $C$ is chosen such that $|S_m| \leq 2C =  n^{\frac{1}{2} + 0.001}$ and let $I = \{ i | x_i \not\in S_i \}$, then 
    \begin{equation*}
        \Pr[|I| \geq (-\ex[D_1] + \varepsilon) n] \leq \exp(-\Omega(n^{0.002})).
    \end{equation*}
\end{prop}
\begin{IEEEproof}
As in the proof of \Cref{prop::del-ch}, we observe that for any $m\in [\varepsilon n, n]$, $\Pr[x_m \not\in S_m | x_m \textup{ was transmitted}]= \Pr[|D_m - \ex[D_m]| > C]$. Evidently, $D_m$ can be seen as the sum of $m$ i.i.d. random variables, each distributed identically to $D_1$ and thus bounded to $[-1,B]$, where $B$ denotes the maximum length of a burst of insertions. By applying Hoeffding's inequality \cite{mitzenmacherProbabilityComputingRandomized2005} and $2C = n^{\frac{1}{2}+0.001}$, we arrive at
\begin{IEEEeqnarray*}{+rCl+x*}
    \Pr[x_m \not\in S_m | x_m \textup{ was transmitted}] &\leq& 2\exp\Big(-\frac{n^{0.002}}{2(B+1)^2}\Big).
\end{IEEEeqnarray*}
 Once again, the union bound lets us deduce that the probability that there exists an $i\in [\varepsilon n, n]$ such that $x_i$ was transmitted and $x_i\notin S_i$ is at most $n \cdot \exp(-\Omega(n^{0.002})) = \exp(-\Omega(n^{0.002}))$.

Since $\Pr[|\bfy| < (1 + \ex[D_1] - \varepsilon) n] = \Pr[D_m -\ex[D_m] \leq -n\varepsilon]$, Hoeffding's inequality again brings us to
\begin{IEEEeqnarray*}{+rCl+x*}
 \Pr[|\bfy| < (1 +\ex[D_1] - \varepsilon) n] 
 &=& \Pr[D_n - \ex[D_n] <  - \varepsilon n]  \\
 &\leq& \exp\Big( -\frac{2\varepsilon^2 n}{(B+1)^2}\Big) \leq e^{-\Omega(n)}.
\end{IEEEeqnarray*}

Lastly, we bound the probability that $|I|\geq (-\ex[D_1] + \varepsilon)n$ by the sum of the probability that $|\bfy| < (1 + \ex[D_1] - \varepsilon) n$ and the probability that given that $|\bfy| \geq (1 + \ex[D_1] - \varepsilon)n$, we have an $i\in [n]$ such that $x_i$ was transmitted and $x_i \notin S_i$. 
The proves the proposition.
\end{IEEEproof}

Analogous to \Cref{prop::del-ch}, this proposition implies that any list-recoverable code may be used over the Davey-MacKay channel.
\begin{thm} \label{thm::davey-ch}
    Let $\pin, \pd\in (0,1)$ and let $\varepsilon > 0$.
    Let $\cC$ be an $[n,k]_q$ code that is $(-\ex[D_1] + \varepsilon, n^{1/2 + 0.001}, L)$ list-recoverable with an efficient list recoverable algorithm $\mathtt{List}$-$\mathtt{recover}$. 
    Let $\bfc\in \cC$ be any codeword and denote by $\bfy$ the output of the Davey-MacKay channel with probabilities $\pin$ and $\pd$ (with a finite insertion burst length) when given $\bfc$ as input.
    
    Let $S_1, \ldots, S_n$ be the lists generated from $\bfy$ according to \Cref{prop::davey-ch}. Then, with probability $\exp(-\Omega(n^{0.002}))$, running $\mathtt{List}$-$\mathtt{recover}$ produces a list $\cL$ such that $\bfc \in \cL$. 

    Moreover, if $(-\ex[D_1] + \varepsilon)n \leq \floor{\frac{\ded(\cC) - 1}{2}}$, then $|\cL| = 1$.
\end{thm}
\begin{remark} \label{rem::davey-rate-tradeoff}
    The rate-error tradeoff is similar to the discussion in \Cref{rem::del-rate-tradeoff}.
\end{remark}


\section{Koetter-Vardy Algorithm for Insdels}

In this section, we adapt a decoding algorithm for Reed-Solomon codes that generalizes list decoding and list recovery, namely the Koetter-Vardy (KV) algorithm \cite{koetterAlgebraicSoftdecisionDecoding2003}, to correct insdel errors. This algorithm extends the Guruswami-Sudan (GS) algorithm of RS codes by incorporating soft information concerning the received symbols. It accomplishes this by using the reliability information of these symbols to form a multiplicity matrix that assigns unequal weights to the interpolation points of the RS code.

\subsection{Reliability Matrix for Insdels}

The primary obstacle to adapting the KV algorithm for use over channels susceptible to insertions and deletions alongside substitutions lies in the computation of the reliability matrix. Since such channels are not memoryless, we rely on the concept of the generalized reliability matrix \cite[Eq. (46)]{koetterAlgebraicSoftdecisionDecoding2003}. Specifically, if we denote a fixed ordering of $\Fq$ as $\alpha_1, \ldots, \alpha_q$, then the $(i,j)$th entry of the reliability matrix, denoted as $\Pi$, should quantify the probability that the $i$th symbol in the transmitted codeword $\bfx \in \Fq^n$ equals $\alpha_j$, for all $j \in [q]$ and $i \in [n]$, given a specific received sequence, say $\bfy$, i.e., $\pi_{i,j} \triangleq \Pr[\bfx_i = \alpha_j | \bfy]$. To compute this quantity, we first note that in the context of an insertion-deletion channel, any transmitted symbol $x_i$ either undergoes transmission with a specific drift, or deletion. Thus, given an $[n,k]_q$ RS code $\cC$, 
and a vector $\bfy \in \Fq^r$ that results from the transmission of some $\bfx\in\cC$ over the Davey-MacKay channel, $\Pr[x_i = \alpha_j | \bfy]$ can be expanded as follows.
\begin{IEEEeqnarray*}{+rCl+x*}
    \Pr[x_i = \alpha_j | \bfy] &=& \frac{ \sum_{\substack{\bfx \in \cC \\ x_i = \alpha_j}} \Pr[\bfy, \bfx]}{ \sum_{\substack{\bfx \in \cC}} \Pr[\bfy, \bfx]} \approx \frac{ \sum_{\substack{\bfx \in \Fq^n \\ x_i = \alpha_j}} \Pr[\bfy, \bfx]}{ \sum_{\substack{\bfx \in \Fq^n}} \Pr[\bfy, \bfx]}. 
\end{IEEEeqnarray*}

The approximation is necessary since marginalizing $\Pr[\bfy, \bfx]$ over (a subset of) $\cC$ is computationally expensive. The quantity $\sum_{\substack{\bfx \in \Fq^n}} \Pr[\bfy, \bfx]$ signifies the probability of receiving a specific sequence $\bfy \in \Fq^r$ given that a vector in $\Fq^n$ was transmitted and is constant for all $\bfy$ of a specific length $r$, 
if all transmitted sequences are equally likely. For brevity of notation, we denote this by $P_n(\bfy)$ and can compute it as the sum of weights of all paths on a two-dimensional lattice from coordinates $(0,0)$ to $(n,r)$, where paths consist of horizontal, vertical and diagonal steps with weights $\pin / q$, $\pd$ and $\pt / q$ respectively \cite{daveyReliableCommunicationChannels2001, banerjeeSequentialDecodingMultiple2024}. Note that horizontal edges are absent in the final row, as insertions cannot follow the transmission of the final symbol. This process accounts for every possible sequence of insertion, deletion, and transmission events, thereby emulating the Davey-MacKay channel depicted in Figure 1. Next, we recall that $x_i$ is either deleted or transmitted with a certain drift, and proceed by decomposing $\sum_{\substack{\bfx \in \Fq^n, x_i = \alpha_j}} \Pr[\bfy, \bfx]$ as follows.
\begin{IEEEeqnarray*}{+rCl+x*}
    &&\sum_{\substack{\bfx \in \Fq^n, x_i = \alpha_j}} \Pr[\bfy, \bfx, x_i \text{ deleted}] \\
    &=& \sum_{l=1}^{r} \Big(\sum_{\substack{\bfx \in \Fq^{i-1}}} \Pr[\bfy_1^{l-1}, \bfx]\Big) \pd \Big(\sum_{\substack{\bfx \in \Fq^{n-i}}} \Pr[\bfy_{l}^{r}, \bfx]\Big), \\
    &&\sum_{\substack{\bfx \in \Fq^n, x_i = \alpha_j}} \Pr[\bfy, \bfx, x_i \text{ transmitted}] \\
    &=& \sum_{l=1}^{r} \Big(\sum_{\substack{\bfx \in \Fq^{i-1}}} \Pr[\bfy_1^{l-1}, \bfx]\Big) \pt \phi(y_l | \alpha_j) \Big(\sum_{\substack{\bfx \in \Fq^{n-i}}} \Pr[\bfy_{l+1}^{r}, \bfx]\Big),
\end{IEEEeqnarray*}
where $\phi(\alpha|\beta) = 1$ if $\alpha=\beta$ and $0$ otherwise. These equations allow us to write
\begin{IEEEeqnarray}{rCl}
\pi_{i,j} &=& \frac{\pd \sum_{l=1}^{r} P_{i-1}(\bfy_1^{l-1})P_{n-i}(\bfy_{l}^{r})}{P_N(\bfy)} \nonumber\\
&& + \frac{\pt \sum_{l=1}^{r} \phi(y_l|\alpha_j) P_{i-1}(\bfy_1^{l-1})P_{n-i}(\bfy_{l+1}^{r})}{P_n(\bfy)}. \label{eq::pi_ij}
\end{IEEEeqnarray}

In case of multiple received sequences, say $\bfy^{(1)}, \ldots, \bfy^{(M)}$, which result from independent transmissions of the encoding of the same information vector, say $\bfu \in \Fq^k$, where $k<n$, the reliability matrix can be computed in a similar spirit as before, i.e.,  $\pi_{i,j} \triangleq {\Pr[\bfx_i = \alpha_j | \bfy^{(1)}, \ldots, \bfy^{(M)}]}$. Note that
\begin{IEEEeqnarray*}{+rCl+x*}
    \Pr[x_i = \alpha_j | \bfy^{(1)}, \ldots, \bfy^{(M)}]
   \!\! &=& \!\! \frac{\sum_{\substack{\bfu \in \Fq^k \\: x_i = \alpha_j}} \Pr[\bfu, \bfy^{(1)}, \ldots, \bfy^{(M)}] }{\sum_{\bfu \in \Fq^k} \Pr[\bfu, \bfy^{(1)}, \ldots, \bfy^{(M)}]} \\
    &\overset{(i)}{=}& \!\! \frac{\sum_{\substack{\bfu \in \Fq^k \\: x_i = \alpha_j}}\! \Pr[\bfu] \prod_{h=1}^{M} \! \Pr[\bfy^{(h)} | \bfu] }{\sum_{\bfu \in \Fq^k} \! \Pr[\bfu] \prod_{h=1}^{M} \Pr[\bfy^{(h)} | \bfu]} \\
    &\overset{(ii)}{=}& \frac{\prod_{h=1}^{M} \sum_{\substack{\bfu \in \Fq^k \\ : x_i = \alpha_j}} \Pr[\bfu, \bfy^{(h)}] }{ \prod_{h=1}^{M} \sum_{\bfu \in \Fq^k} \Pr[\bfu, \bfy^{(h)}]} \\
    &=& \prod_{h=1}^{M} \frac{\sum_{\substack{\bfx \in \cC, x_i = \alpha_j}} \Pr[\bfx, \bfy^{(h)}] }{ \sum_{\substack{\bfx \in \cC}} \Pr[\bfx, \bfy^{(h)}]} \\
    &=& \prod_{h=1}^{M} \Pr[x_i = \alpha_j | \bfy^{(h)}],
\end{IEEEeqnarray*}
where $(i)$ follows from the fact that the received vectors are acquired from independent transmissions of the same codeword and $(ii)$ results from uniform distribution of all information vectors.

Observe that the two-dimensional lattice used to compute $P_n(\bfy_1^r)$, as discussed earlier, can also be flipped along the off-diagonal, enabling the traversal of the finite state machine in reverse, i.e., from the ending to the starting state. Hence, the order of received and transmitted symbols is altered: the columns, from left to right, correspond to the symbols \( y_r, \ldots, y_1 \), while the rows, from top to bottom, represent the symbols \( x_n, \ldots, x_1 \). This change may impact the second term in (\ref{eq::pi_ij}) due to asymmetry introduced by the non-uniform diagonal edge weights \( P_t \phi(y_r | \alpha_j) \). For instance, when using the forward lattice, one or more symbols in \( \bfy^{r-1}_1 \) affect \( \pi_{n,j} \), but this is not the case in the reversed model. To remedy this, we use a bidirectional version of the computation of $\pi_{i,j}$ presented in (\ref{eq::pi_ij}), i.e., we compute $\pi_{i,j}$ on the forward and the reversed lattice, and ultimately use the average of the two.

\subsection{Numerical Results}

    \begin{figure}[t]
		\centering
        \scalebox{1.05}{
		\begin{tikzpicture}
			\begin{axis}[legend style={nodes={scale=0.65, transform shape}},
				xmode=log, ymode=log,
				xmin=8e-4, xmax=0.12,
				ymin=2e-4, ymax=1.0,
				xlabel=\textsc{$P$}, ylabel={FER},
				legend pos=south west, legend style={font=\scriptsize},
				grid=both]
                \addplot plot[solid, green!20!black, mark=diamond*, mark options={fill=green!20!black}] coordinates {
                    (0.004, 0.0177522)
                    (0.008, 0.057011) 
                    (0.016, 0.164)
                    (0.03, 0.3508)
				};
				\addlegendentry{$M=1$, $\pd = \pin = P/2$}
                \addplot plot[solid, red, mark=diamond*, mark options={fill=red}] coordinates {
                    (0.004, 0.01169)
                    (0.008, 0.042)
                    (0.01, 0.0592)
                    (0.015, 0.13105) 
                    (0.02, 0.2142)
				};
				\addlegendentry{$M=1$, $\pd = P$, $\pin = 0$}
                \addplot plot[solid, blue, mark=diamond*, mark options={fill=blue}] coordinates {
                    (0.004, 0.0104) 
                    (0.008, 0.0405)
                    (0.01, 0.063)
                    (0.015, 0.1258) 
                    (0.02, 0.2124)
				};
				\addlegendentry{$M=1$, $\pin = P$, $\pd = 0$}
                \addplot plot[dashed, green!20!black, mark=square*, mark options={fill=green!20!black}] coordinates {
                    (0.008, 0.015)
                    (0.016, 0.056)
                    (0.02, 0.0826)
                    (0.03, 0.1467) 
				};
				\addlegendentry{$M=2$, $\pd = \pin = P/2$}
                \addplot plot[dashed, red, mark=square*, mark options={fill=red}] coordinates {
                    (0.004, 0.0014)
                    (0.008, 0.0042)
                    (0.01, 0.00608)
                    (0.015, 0.02109)
                    (0.02, 0.03476) 
				};
				\addlegendentry{$M=2$, $\pd = P$, $\pin = 0$}
                \addplot plot[dashed, blue, mark=square*, mark options={fill=blue}] coordinates {
                    (0.008, 0.00218)
                    (0.01, 0.0047)
                    (0.015, 0.0125)
                    (0.02, 0.0252) 
                    (0.03, 0.0569)
				};
				\addlegendentry{$M=2$, $\pin = P$, $\pd = 0$}
                \addplot plot[dash dot, green!20!black, mark=*, mark options={fill=green!20!black}] coordinates {
                    (0.008, 0.001125)
                    (0.016, 0.0128)
                    (0.02, 0.021184)
                    (0.04, 0.08578)
				};
				\addlegendentry{$M=4$, $\pd = \pin = P/2$}
                \addplot plot[dash dot, red, mark=*, mark options={fill=red}] coordinates {
                    (0.015, 0.001641137)
                    (0.02, 0.00383) 
                    (0.03, 0.013829) 
				};
				\addlegendentry{$M=4$, $\pd = P$, $\pin = 0$}
                \addplot plot[dash dot, blue, mark=*, mark options={fill=blue}] coordinates {
                    (0.015, 0.0004) 
                    (0.02, 0.0016003)
                    (0.03, 0.0087)
				};
				\addlegendentry{$M=4$, $\pin = P$, $\pd = 0$}
			\end{axis}
		\end{tikzpicture}}
  \vspace{-2ex}
		\caption{Frame error rates for a $[100,33]$ RS code over field $\mathbb{F}_{101}$, where $M$ denotes the number of received sequences, $\pd$ the deletion probability and $\pin$ the insertion probability. 
        }
		\label{fig::fer}
	\end{figure}
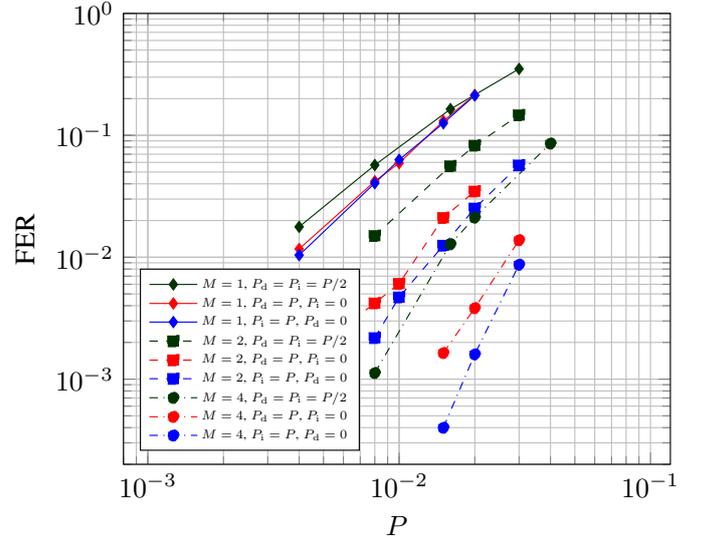

To assess the performance of the proposed decoder for correcting insdels, we use a primitive $[100,33]$ RS code over the field $\mathbb{F}_{101}$, with its evaluation points randomly permuted. A random permutation of the evaluation points is chosen because, according to \cite{beelenReedSolomonCodesInsertions2025}, the respective RS code is more likely to have a better insdel correction capability. We set the list size parameter to $5$ and simulate the transmission of its codewords over channels with varying insertion and deletion probabilities and different numbers of received sequences. No limit on the maximum length of a burst of insertions is imposed. To limit computational complexity, the marginalization in the numerator of $\pi_{i,j}$ in (\ref{eq::pi_ij}), is performed over a smaller set of indices centered around the expected number of received symbols after the transmission of $x_i$. 
The frame error rates observed are shown in \Cref{fig::fer}. As expected, the error rates decrease with an increasing number of received sequences. It also seems that correcting deletions is more challenging than correcting insertions. This can be attributed to the fact that insertions only shift the transmitted symbols further to the right, allowing the decoder to still access the originally transmitted symbols, albeit with a potentially lower multiplicity. In contrast, deletions result in the complete loss of transmitted symbols. A combination of insertion and deletion errors is even more difficult to correct, as it becomes harder to ascertain the true positions of the transmitted symbols in the received sequence.


\balance
\printbibliography

\end{document}